\documentstyle[multicol,aps,epsfig]{revtex}

\begin{document}
\title{Pattern Dynamics of Vortex Ripples in Sand: \\
  Nonlinear Modeling and Experimental Validation}
\author{K.H.~Andersen$^1$, M.~Abel$^2$, J.~Krug$^3$, C.~Ellegaard$^4$,
  L.R.~S{\o}ndergaard$^{4,5}$ and J.~Udesen$^{4,5}$}
\address{$^1$Department of Mechanical Engineering, Technical
  University of Denmark, DK-2800 Kgs.~Lyngby} 
\address{$^2$Institut
  f{\"u}r Physik, Universit{\"a}t Potsdam, D-14415 Potsdam, Germany}
\address{$^3$Fachbereich Physik, Universit{\"a}t Essen, D-45117 Essen,
  Germany.}  
\address{$^4$Niels Bohr Institute, University of
  Copenhagen, DK-2100 Copenhagen {\O}, Denmark.}
\address{$^5$Department of Mathematics and Physics, University of
  Roskilde, Box 260, DK-4000 Roskilde, Denmark.}  
\date{\today}
\maketitle

\begin{abstract}
  Vortex ripples in sand are studied experimentally in a
  one-dimensional setup with periodic boundary conditions. The
  nonlinear evolution, far from the onset of instability, is analyzed
  in the framework of a simple model developed for homogeneous
  patterns. The interaction function describing the mass transport
  between neighboring ripples is extracted from experimental runs
  using a recently proposed method for data analysis, and the
  predictions of the model are compared to the experiment.  An
  analytic explanation of the wavelength selection mechanism in the
  model is provided, and the width of the stable band of ripples is
  measured.

\end{abstract}
\pacs{PACS numbers\,: 
45.70.Qj, 
47.54.+r, 47.20.Lz
}
\begin{multicols}{2}
%
%
Ever since the establishment of a conceptual framework for pattern formation
\cite{cros:93}, the description of patterns formed in sand
by the flow of wind or water has posed a challenge to the community
\cite{nish:93,wern:93,csah:00,Bet,steg:99,hans:01}. 
Despite diverse efforts including 
coupled map models for ripples and
dunes in air \cite{nish:93}, stochastic models for ripples in air
\cite{wern:93} or continuum equations based on the symmetries of the
problem \cite{csah:00}, theoretical understanding has
remained sparse. For example, all models display a coarsening of the
ripple/dune pattern, but the coarsening does not terminate at a final
selected wavelength, as is frequently observed in nature. 
Furthermore the models are heuristic, and it is not possible to
make a quantitative comparison with experiments.  

Here, we study {\em vortex
ripples} \cite{bagn:46}, which are created by an oscillatory water
flow, such as that generated near the sand bed by a surface wave. Vortex
ripples have attracted attention as an example of a non-linear pattern
forming system with a strongly sub-critical first bifurcation
\cite{steg:99,hans:01,ande:01}, which cannot be described by
conventional methods like amplitude equations \cite{ande:01}. As most
other sand patterns, they display coarsening and saturation at a
finite wavelength. The approach pursued in this Letter combines a
simple model for the fully developed pattern with a sophisticated data
analysis which allows to extract the key model ingredient -- the
interaction function $f(\lambda)$ -- directly from the experimental
runs. In this way the validity of the model can be tested, and
additional features required for the description, can be
identified. As far as we know, this is the first quantitative comparison
between theory and experiment for a sand pattern. The basic ideas are
general, and we expect that the theoretical formalism combined with
the experimental analysis can be used for related sand patters
(e.g. the one studied in \cite{Bet}), or other strongly nonlinear systems.

One dimensional vortex ripples can be created in an annular channel
\cite{steg:99,sche:99}, ensuring the pattern to be subject to
well-defined, periodic boundary conditions.  Freely grown ripples are
created from a flat bed by a coarsening process, which eventually
saturates at an equilibrium state, where the ripple length
$\tilde{\lambda}$ is almost independent of the frequency $\nu$ of the
driving and proportional to the amplitude of the oscillation of the
plate $a$ \cite{bagn:46,niel:81}.  Here we are mostly concerned with
the stability and evolution of the ripple patterns themselves, and not
with the instability of the flat bed, which has been discussed
elsewhere \cite{blon:90,ande:01}.  By creating a
homogeneous ripple pattern, where all ripples have the same
length, and changing amplitude and frequency, the stability
of the pattern is probed.  In this way it was found that there is a
stable band of ripples, $\lambda_{min}
\!<\!\tilde{\lambda}\!<\!\lambda_{max}$, for a given set of driving
conditions \cite{hans:01}.

\begin{figure}[htb]
   \epsfig{file=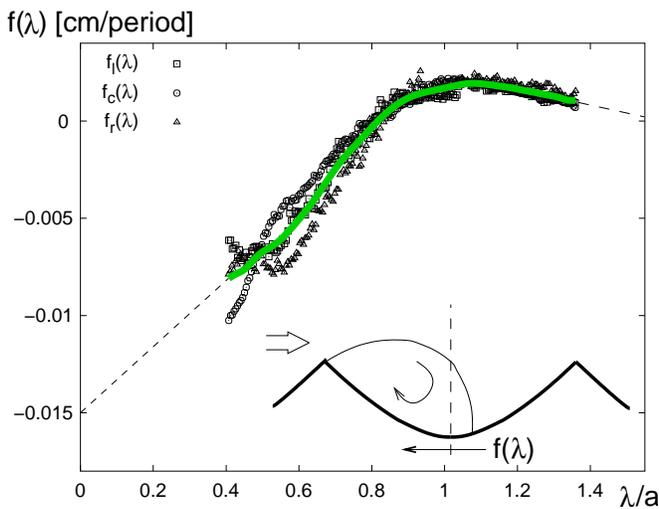,width=\columnwidth}
  \caption{The interaction function for $a\!=\!6$ cm and $\nu\!=\!0.6$
    Hz. The thick gray line is a smooth curve drawn through the points
    representing the three different functions. The right
    limit of the plot at $\lambda/a=1.45$ is the limit of stability,
    where new ripples are created ($\lambda_{max}$). The inset shows a
    sketch of the ripples in the part of the oscillation when the flow
    is from the left to the right. The interaction function can be
    interpreted as the transport of sand in the trough, across the
    dashed line.}
  \label{fig:f}
\end{figure}

Recently, a simple model was proposed, which describes both  the
coarsening and saturation of  ripples, and reproduces the
existence of a stable band \cite{ande:01}.  In its simplest form, the
change of the length $\lambda_j$ of the ripple $j$ is a function of
$\lambda_j$ itself and the lengths of the neighboring ripples,
\begin{equation}
  \dot{\lambda}_j = -f(\lambda_{j-1}) + 2f(\lambda_j) -
  f(\lambda_{j+1}),
  \label{equ:model}
\end{equation}
where the {\em interaction function} $f(\lambda)$ describes the
transfer of mass between neighbouring ripples (Fig.~\ref{fig:f}).
Arguments based on the properties of the separation vortex forming
behind a ripple show that $f(\lambda)$ should be convex,  its
approximate shape has been found from a numerical simulation of the
fluid and sand flow over homogeneous ripples \cite{ande:01}.  The
 purpose of this Letter is a  test of the model
(\ref{equ:model}) through comparison with experimental
data. 

\begin{figure}[htb]
  \epsfig{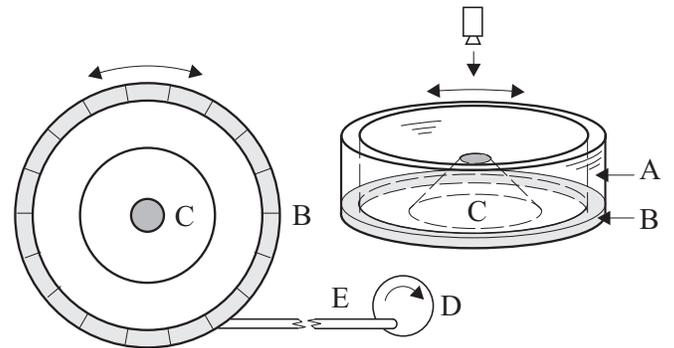}
  \caption{A sketch of the experimental setup seen from above (left)
    and from the side (right). The length of the arm E and the width
    of the channel are not to scale.}
  \label{fig:experiment}
\end{figure}

\paragraph*{Experimental setup.}
A 11 mm wide, 15 cm high annular plexiglass channel of diameter 48.6 cm
is filled with water, A, and glass beads, B, with a diameter of 250
$\pm 50\ \mu$m (Fig.~\ref{fig:experiment}). In the middle is a conical
mirror, C, which is filmed from above by a CCD camera with a
resolution of 640$\times$480 pixels. The channel is attached to a
motor, D, by a 1.5 m long arm E, which oscillates the channel in an
almost sinusoidal fashion. 

\begin{figure}[htb]
  \noindent\epsfig{file=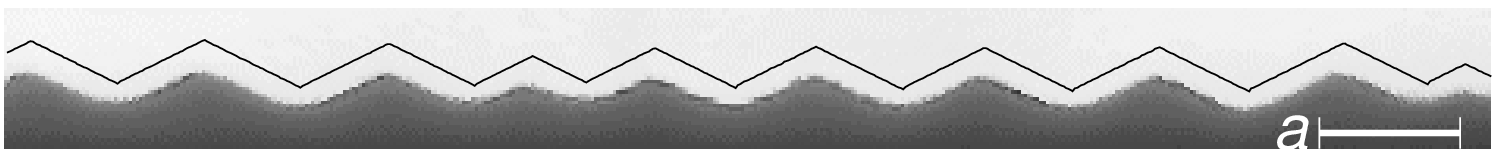,width=\columnwidth}
  \epsfig{file=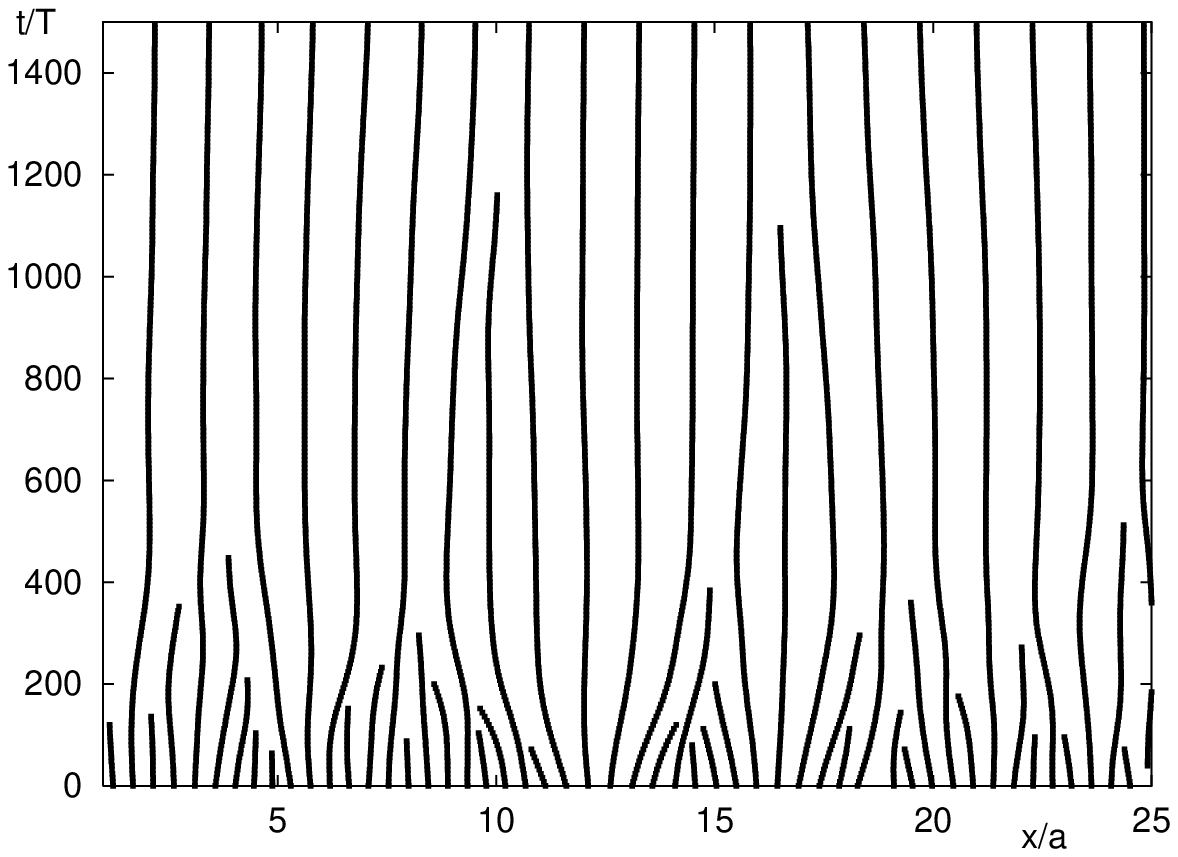,width=\columnwidth}
  \epsfig{file=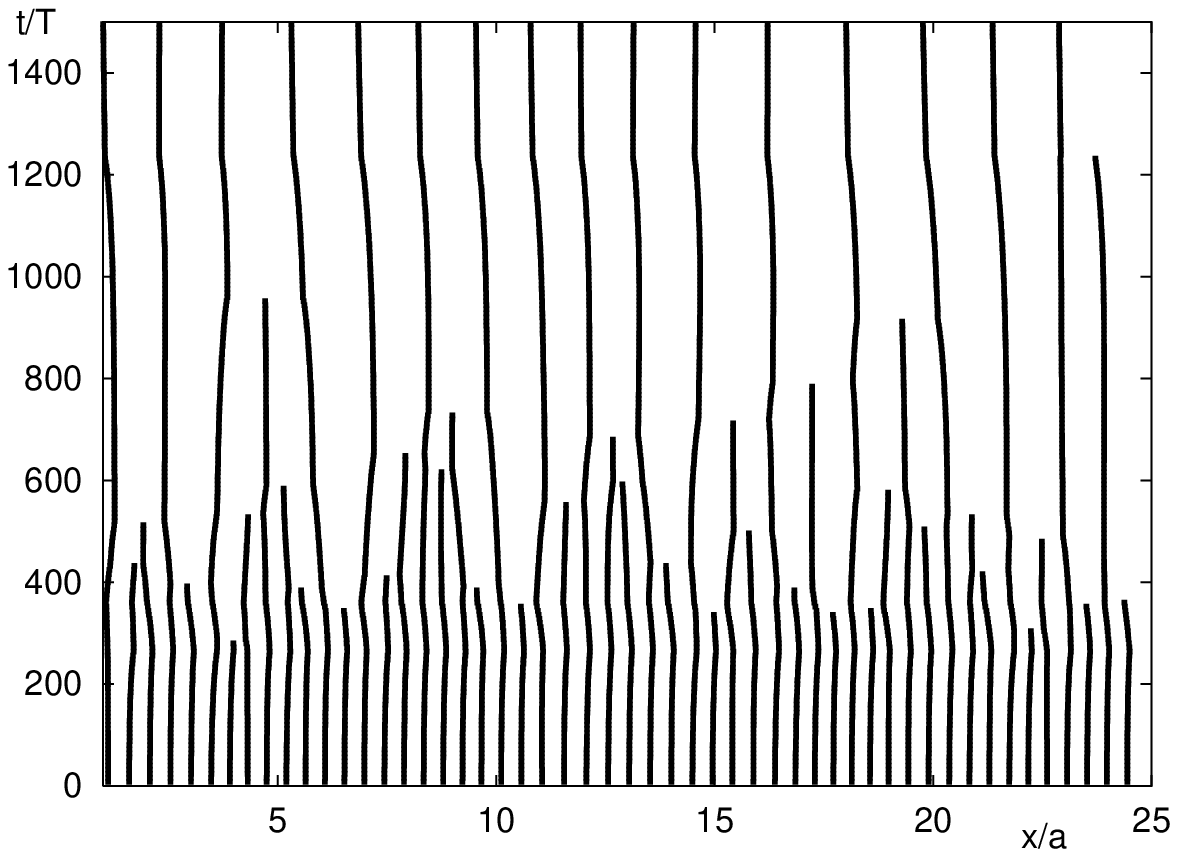,width=\columnwidth}
  \caption{a) An example of an extracted profile (dashed line) and the
    fitted triangles with constant slope. The line is shown above the
    profile for clarity. b) The experimental evolution of the position
    of the ripple crests starting from ripples with lengths $2.5$ cm
    and evolving with $a\!=\!6$ cm and $\nu\!=\!0.6$ Hz. c) A
    simulation of the model (\protect\ref{equ:model}) using the
    extracted interaction function and the same initial conditions as
    above.}
  \label{fig:big}
\end{figure}

We create initial conditions with small ripples
by oscillating at a small amplitude. The experiment starts when 
amplitude and frequency are changed to their desired values. Here, we
focus on the condition $a\!=\!6$ cm, $\nu\!=\!0.6$ Hz. We want to
emphasize that there is nothing special about this condition, the same
qualitative results are obtained for other parameters. Growing the
ripples from initial conditions with homogeneous ripples smaller than
$\lambda_{min}$
results in a final number of 19--21 ripples with a length (averaged
over 16 realizations) of $\tilde{\lambda}/a\!=\!1.25$. The uncertainty
in the length due to periodicity is around 5~\%.  The images are
transformed back to cartesian coordinates, and the lengths 
$\lambda_j$ are found by fitting them to  triangles with
fixed slope (Fig.~\ref{fig:big}a).  From series of consecutive
$\lambda_j$, the temporal change, $\dot{\lambda}_j$, is
established, and a space-time plot of the ripple evolution
is constructed. Fig.~\ref{fig:big}b shows the evolution starting
from a small initial wavelength, which leads to the annihilation of
ripples.  Around each annihilation the ripples have not been analysed,
as it becomes inaccurate to fit the triangles during these events.

\paragraph*{Measuring the stable band.}
By running the experiment with different values of $a$ until a
homogeneous pattern occurs, we create initial conditions with
different no.~of ripples. From this initial pattern, the experiment is
started with the above parameters, and run for 10000 periods. We have
created initial conditions with $N\in[17;24]$ and observed that for
$N>22$ ripples are annihilated, while for $N<18$ one or more new
ripples are created.  We therefore conclude that there exists a stable
band $N\in[17.5;22.5]$ (for $a=6$ cm, $\nu =0.6$ Hz), which
corresponds to $\lambda_{min}/a=1.13 \pm 0.03$ and $\lambda_{max}/a =
1.45 \pm 0.05$.  Note that this is a more accurate measurement of the
stable band than that conducted in \cite{steg:99,hans:01}, as we are
always forcing with the same values of $a$ and $\nu$.

\paragraph*{Mass transfer model.}
The model (\ref{equ:model}) was originally presented in
Ref.~\cite{ande:01}, together with a more refined version.  A
stability analysis of the homogeneous state $\lambda_j \equiv \bar
\lambda$ shows that ripples are stable (unstable) if $f'(\bar \lambda)
< 0$ ($> 0$).  For a convex interaction function, the lower stability
boundary $\lambda_{min}$ therefore lies at the maximum
$\lambda_{marg}$ where $f'(\lambda_{marg}) = 0$.  To investigate
nonlinear pattern evolution within this model, Eq.~(\ref{equ:model})
is supplemented by the rule that ripples which reach zero length are
annihilated and removed from the system of equations, while the
remaining ones are relabeled.

Simulations of the model starting from typical initial conditions
\cite{note:fronts} in the unstable band $\lambda < \lambda_{marg}$
show that an equilibrium wavelength is reached which is essentially
independent of the initial wavelength, and depends only on the
interaction function $f(\lambda)$.
To gain some insight into the selection mechanism,
it is useful to recast the model into potential form by writing it in
terms of the position of the troughs between the ripples $x_j$ defined
by $\lambda_j = x_{j+1}-x_j$. Then $\dot{x}_j = -\partial V/\partial x_j$,
with the potential $V$ given by
\begin{equation}
 V = - \sum_{j=1}^N \int_0^{x_{j+1}-x_j} d\lambda \,f(\lambda) ,
\end{equation}
where $N$ is the number of ripples. It is then plausible to conjecture
that the equilibrium length $\tilde{\lambda}$ can be found by
minimizing $V$ for homogeneous ripples, under the constraint that the
total length $L=N\tilde{\lambda}$ is conserved. This implies that
$\tilde \lambda$ is determined through the Maxwell construction
applied to $f$: 
\begin{equation} 
\label{lever}
\int_0^{\tilde{\lambda}} d\lambda \; f(\lambda) = 
\tilde{\lambda} f(\tilde{\lambda}).
\end{equation}
Comparison with numerical simulations shows that (\ref{lever})
systematically overestimates $\tilde{\lambda}$, with a better
performance the steeper the stable branch of $f$.  Our interpretation
is that the deterministic dynamics gets stuck in the multitude of
metastable states of (\ref{equ:model}); recall that \emph{any}
homogeneous state with $\bar \lambda > \lambda_{marg}$ is a stable,
stationary solution.
Since the mean ripple length can increase only by annihilations, its
evolution freezes once all ripples are in the stable band.  The
wavelength predicted by (\ref{lever}) should therefore be an upper
bound on the actual equilibrium wavelength, which is true in all cases
we have considered.


\paragraph*{Data analysis.}
Given the time series $\lambda_j(t)$ we want to (i) evaluate 
how well the model (\ref{equ:model}) describes the evolution of the ripples
 and (ii) extract the interaction function
$f(\lambda)$. For the analysis, we write (\ref{equ:model})
in the more general form
\begin{equation}
  \dot{\lambda}_j = -f_l(\lambda_{j-1}) + 2f_c(\lambda_j) -
  f_r(\lambda_{j+1})\;, 
  \label{equ:analysis}
\end{equation}
with an additional degree of freedom as  the functions are not
required to be equal.

We want to determine the optimal transformations
$f_l(\lambda_{j-1}),\,f_c(\lambda_{j}),\,f_r(\lambda_{j+1})$ in the
sense that they minimize the error $\chi^2= \sum_t \left[
\dot{\lambda}_j(t) + f_l(t) - 2f_c(t) + f_r(t)\right]$.  The problem
is solved numerically by the ACE algorithm
\cite{Breiman-Friedman-85,Voss-Kolodner-Abel-Kurths-99}. The algorithm
works by varying $f_{l,c,r}$ in the space of all measurable functions
until convergence to the {\it absolute} minimum of $\chi^2$
\cite{Breiman-Friedman-85}.  The results are nonparametric functions
which are given in numerical form by points, e.g., ($\lambda_j,
f_c(\lambda_j)$).  An upper bound on the error, e.g., on $f_c$,
is given by $\sigma_{f_c} = \max_{\lambda}(df_c/d\lambda)\cdot
\sigma_\lambda$, with $\sigma_\lambda$ being the measurement error at
the point $\lambda$; analogous estimates apply to f $f_l,f_r$.
The error in the points on the $\lambda$-axis is
estimated using the above result and equals roughly the errors in the
given values of $\lambda$.

The quality of each of the resulting functions is given by the maximum
correlation $\psi_{l,c,r}$ of one of the terms with the sum of all the
others.
A value of $\psi=1$ implies a perfect result, lower values indicate
either imperfect modeling or (measurement) noise or both.

\paragraph*{Results.}
We have performed the analysis on data from 18 realizations of a
coarsening process initiated with approximately 60 small ripples and
run with $a=6$ cm and $\nu=0.6$ Hz.  Fig.~\ref{fig:f} shows the three
functions $f_l$, $f_c$ and $f_r$ averaged over the 18 runs.  Clearly
there is some noise, but the functions are very similar as expected
from (\ref{equ:model}). The maximum correlations are found to be
$\Psi_l=0.66$, $\Psi_c=0.88$, $\Psi_r=0.72$.  The uncertainty in
estimating $\lambda$ amounts to about 2 mm, with a maximum slope of
$f$ of around $3\times 10^{-3}$ (period)$^{-1}$, the absolute error of
$f$ is $6\times 10^{-4}$ cm/period. This means that the result does
not fluctuate due to lack of data, rather the model
(\ref{equ:analysis}) does not account for all the variation in the
ripple evolution.  We will return to this point later.


The maximum of the function lies around $\lambda_{marg}\!=\!1.08a$,
which is a little larger than what was found by the numerical
simulations in \cite{ande:01}.  Concerning the shape of the function,
it is interesting that the slope in the unstable band $\lambda <
\lambda_{marg}$ is much larger than the slope in the stable band
$\lambda > \lambda_{marg}$. As the slope is a measure of the time
scale of the dynamics, this implies that the initial coarsening stage
is much faster than the equilibrating stage.

To use the result function to integrate (\ref{equ:model}) numerically for
evaluating (\ref{lever}), information about the smallest ripples is
needed. In lack of such information, we have extrapolated linearly to 
$\lambda = 0$, Fig.~\ref{fig:f}, dashed line.
Varying the slope of the extrapolated part by a factor of two in
either direction does not change the final number of ripples.
Fig.~\ref{fig:big}c shows a space-time plot of the numerical integration.
The model predicts a
qualitatively similar behaviour as the experiment, however the
instability of the small ripples develops slower  in the
model than in the experiment. If the extrapolated slope for the
interaction function is made steeper, this instability will evolve
faster in the model. The model also overestimates the final ripple length. 
For similar initial conditions and system size, the final
no.~of ripples is typically 18, whereas the experiment yields 20.

Evaluating the Maxwell construction (\ref{lever}) using the measured function
and extrapolating to account for the larger ripples, gives
$\tilde{\lambda} = 2.2a$, which lies far outside of the range of
definition of $f(\lambda)$.  Evidently, the experimentally determined
$f(\lambda)$ belongs to the class of interaction functions for which
the analytic bound is not useful.

The upper bound of the stable band, where new ripples are created, was
found to be $\lambda_{max} = 1.45a$. At this point, an infinitesimally
small ripple, inserted in the trough between two larger ripples, is
able to gain mass from the neighbours, and thus grows. In the model,
this corresponds to ripple lengths for which the interaction function
is smaller than the value of the interaction function for ripples at
zero length, i.e., $f(\lambda_{max}) = f(0)$ \cite{ande:01}.
Returning to the measured interaction function in Fig.~\ref{fig:f}, we
see that the interaction function at $\lambda_{max}$ is approximately
zero, and at $\lambda=0$ it is extrapolated to $-0.015$ cm/period.
Even with the large uncertainty inherent in the extrapolation, this
does not fit with the theoretical picture given above. We can
therefore conclude, that the model in the form (\ref{equ:model}), is
not able to quantitatively predict creation of new ripples. The reason
for this apparent failure relates to the model being developed
essentially as an expansion around a homogeneous state \cite{ande:01}.
But in the case of a creation, the state is as inhomogeneous as it can
be: a very tiny ripple flanked by large ripples. To account for this,
the interaction function should be described as a function of the
length of the ripple creating the separation bubble, $\lambda_i$, {\it
  but also} of the length of its neighbours.  Ongoing numerical work
indicates that it is most relevant to write $f \equiv f(\lambda_i,
\lambda_{i+1})$where $\lambda_i$ is the length of the ripples creating
the separation bubble, and $\lambda_{i+1}$ is the one ``touched'' by
the separation bubble (see inset Fig.~\ref{fig:f}). In fact, the whole
coarsening process is dominated by highly inhomogeneous
configurations, and this might also be why simulating
(\ref{equ:model}), using the measured interaction function, does not
produce exactly the correct final ripple length.

\paragraph*{Conclusions.}
We have demonstrated that we can extract the interaction function
$f(\lambda)$ from spatio-temporal data of the evolution of the profile
of the sand surface. The nonlinear data analysis shows its full
strength, producing nonparametric function estimates, where any
parametric approach would have failed.  With the interaction function
it is possible to model the evolution of the ripples, in particular
their coarsening and equilibration.

The prediction from the model of the existence of a stable band is
indeed observed in the experiment. The lower bound is predicted at
$\lambda_{marg}/a = \!1.08 \pm 0.03$, while the measurement gives
$\lambda_{min}/a = 1.13 \pm 0.03$.
In a similar experiment \cite{steg:99}, the lower bound of the stable
band was found to coincide with the final ripple length:
$\lambda_{min}=\tilde{\lambda}$, whereas we find that
$\lambda_{min}\!<\!\tilde{\lambda}$. The difference between the two
results might only be apparent, as the number of ripples in the
annulus of \cite{steg:99} was 10 or smaller, giving rise to an
uncertainty in the ripple length on the order of the difference
between $\lambda_{min}$ and $\tilde{\lambda}$.

The dynamics responsible for the evolution of a 1d ripple pattern,
are also relevant for 2d ripple patterns. However, in the latter case,
topological defects will be present \cite{hans:01}, and
the final length selection might be determined by the motion of
these \cite{ande:01}.



\paragraph*{Acknowledgements}
It is a pleasure to acknowledge discussions with T.~Bohr, M.~van Hecke
and F.~Schmidt. J.K. is grateful to DTU and NBI for gracious
hospitality, and to DFG within SFB237 for support. K.H.A thanks
Universit\"at Essen for hospitality.



\end{multicols}

\end{document}